\documentclass[10pt,times,epsfig,psfig,twocolumn,algorithm,algorithmic]{IEEEtran}
\def\BibTeX{{\rm B\kern-.05em{\sc i\kern-.025em b}\kern-.08em
    T\kern-.1667em\lower.7ex\hbox{E}\kern-.125emX}}
\usepackage{epsfig}
\usepackage{amsmath}
\usepackage{algorithm}
\usepackage{algorithmic}

\begin{document}

\title{uRbAn: A M{\em u}ltipath {\em R}outing {\em b}ased {\em A}rchitecture with E{\em n}ergy and Mobility Management for Quality of Service Support in Mobile Ad hoc Networks}
\author{Ash Mohammad Abbas\\
Department of Computer Engineering\\
Aligarh Muslim University\\
Aligarh - 202002, India.}

\maketitle
\thispagestyle{empty}\pagestyle{empty}

\begin{abstract}
Designing a wireless node that supports {\em quality of service} (QoS) in a mobile ad hoc network is a challenging task. In this paper, we propose an architecture of a wireless node that may be used to form a mobile ad hoc network that supports QoS. We discuss the core functionalities required for such a node and how those functionalities can be incorporated. A feature of our architecture is that the node has the ability to utilize multiple paths, if available, for the provision of QoS. However, in the absence of multiple paths it can utilize the resources provided by {\em a single} path between the source and the destination. We follow the modular approach where each module is expanded iteratively. We compare the features of our architecture with the existing architectures proposed in the literature.  Our architecture has provisions of energy and mobility management and it can be customized to design a system-on-chip (SoC).   
\end{abstract}

\section{Introduction}
In many situations, it is not possible or is not advisable 
to communicate through a network using an existing infrastructure. For example,  in deep rural areas, there may not be an infrastructure available to facilitate the communication. On the other hand, the members of a spying team want to communicate without using the existing infrastructure due to security reasons. An ad hoc network can be formed instantaneously enabling the participants to communicate without the required intervention of a centralized infrastructure or an access point. In fact, those who want to bypass the infrastructure for some reasons and still want to communicate may form an ad hoc network. An ad hoc network may provide cheaper and cost-effective ways to share information among many mobile hosts. Some of the characteristics of an ad hoc network differentiate it from other classes of wired and wireless networks. In an ad hoc network, the transmission range of mobile devices is limited, therefore, routes are often multihop. There are no separate routers, therefore, nodes in the network cooperate to forward packets of one another towards their ultimate destinations. The devices used to form such a network are often powered through batteries whose power depletion may cause node failures. Further, nodes may move about randomly and therefore the topology of the network may vary dynamically. 

It is desirable to have a provision of quality of service (QoS) in an ad hoc network. However, there are many peculiar characteristics of an ad hoc network that hinder in providing QoS. For example, the absence of any centralized infrastructure and the dynamically varying topology of a mobile ad hoc network make the provision of quality of QoS a challenging task. Further, the topology information of the network is not available apriory at a central node. A node knows only about its neighbors. As a result, a solution or scheme should be able to work with the localized topology information and in a distributed manner. In other words, devising schemes for providing any hard guarantees about the QoS is difficult due to frequent node and link failures. 

 Many researchers have proposed different architectures that may be used in wireless networks. One such architecture is called {\em network on chip} (NoC) \cite{cidon}. An NoC architecture for provision of QoS using {\em multiprotocol label switching} (MPLS) is proposed in \cite{kim} for networks that require an infrastructure support. Another NoC architecture for heterogeneous traffic support with non-exclusive dual-mode switching is proposed in \cite{secchi} which is suitable for Multi-Processors System on Chip (MPSoC). All these architectures are for wired networks.  

In \cite{zhao}, a Dual-Channel Binary-Countdown Medium Access Control (DBC-MAC) that can be used in {\em wireless network on chip} (WNoC) is proposed. There are two basic components of a WNoC: (i) Transparent Network Interface (TNI), and (ii) Radio Frequency (RF) Node. In general, DBC-MAC resolves contention among RF nodes, and is a cross-layer, synchronized and distributed protocol. An architecture for a wireless node and its implementation for multihop mesh networks in IP-based broadband fixed wireless access systems is proposed in \cite{kishi}. Therein, the wireless node consists of set of outdoor units, an indoor unit, and a network control unit. A multi-channel multi-radio wireless mesh node architecture for {\em network simulator (ns-2)} is proposed in \cite{ji}. A QoS-aware and energy efficient wireless node architecture is proposed in \cite{lettieri}. The architecture uses a router based approach. The architectures proposed in \cite{zhao}, \cite{kishi}, \cite{lettieri}, \cite{ji},   are for wireless nodes that can be used for networks with infrastructures, and they cannot be used in an ad hoc network.   

An architectural framework called Joint Architecture Vision for Low Energy Networking (JAVeLEN) for low energy ad hoc wireless networks is proposed in \cite{redi}. The architecture proposed, therein, employs the design of a dual radio transceiver and requires to redesign of the protocol stack. On the other hand, a wireless sensor node architecture using remote power charging for interactive applications is proposed in \cite{dsouza}.

The design of a node that can be used to form an ad hoc network that supports some form of QoS is a challenging task. There has to be some peculiar functionalities inside a node so as to be used for applications that require a provision of QoS. The major functionalities of a wireless node may include resource management, routing, QoS provisioning, mobility management, energy management, etc. In this paper, we present an architecture for the design of a wireless node for a mobile ad hoc network with QoS support. We discuss the core functionalities required for such a node and how those functionalities can be incorporated. We follow the modular approach where each module is expanded iteratively. Our architecture 
can be customized to design a system-on-chip (SoC) or a system-in-package (SiP).   

The rest of the paper is organized as follows. In Section II, we present outlines of the proposed architecture. In Section III, we describe 
the routing protocol and its components. Section IV contains a description of the components of mobility management. In Section V, we describe the components required for the provision of QoS. Section VI contains components related to the energy management in the wireless node. Section VII contains miscellaneous functions. Finally, the last section is for conclusion.

\begin{figure}
\centerline{\psfig{figure=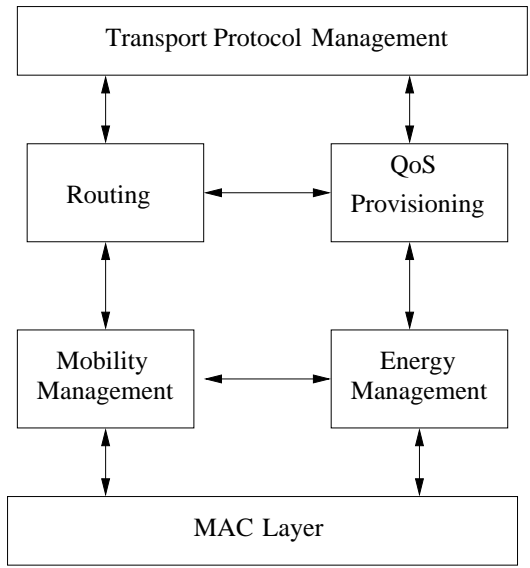, width=3.0in}}
\caption{Major functionalities of a wireless node and their interactions.}
\label{major-func}
\end{figure}
 
\section{Proposed Architecture}
In this section, we describe the proposed architecture of a wireless node. We call the proposed architecture as M$u$ltipath $R$outing $b$ased $A$rchitecture with E$n$ergy and Mobility Management (uRbAn) for Quality of Service Support in mobile ad hoc networks.  The major functionalities that a node should possess are: 
(i) Transport Protocol Management (ii) Routing (iii) QoS Provisioning (iv) Medium Access Control (MAC)  (v) Energy Management (vi) Mobility Management. 
A block diagram depicting these major functionalities is shown in Figure \ref{major-func}. Each of these major functionalities can be further divided into sub-modules. We describe their division as follows.

\begin{itemize}


\item {\em Transport Protocol Management:} This is responsible to transport the packets from one end to another end. The end points are often called sockets. Depending upon the application, this module invokes a transport layer protocol such as TCP, UDP or their variations.

\item {\em Routing:} A routing module consists of a routing protocol that is used to build forwarding tables. A routing protocol may consists of neighbor discovery mechanism, route discovery, route maintenance, and path selection mechanism.

\item {\em QoS Provisioning:} QoS provisioning module contains admission control, priority management of packets, resource management, and scheduling.

\item {\em Medium Access Control:} A MAC provides the functions such as handling access to a shared medium, and allowing multiple devices to uniquely identify one another at data link layer.

\item {\em Energy Management:} This module is responsible for managing the energy or power. For example, the device may enter into sleep state when it is not transmitting or receiving for a timeout. It also manages the transmission and reception powers.

\item {\em Mobility Management:} This module is responsible for actions taken when a device moves. For example, the device need to update its routing and forwarding tables if the links to its current neighbors are broken due to mobility and if it detects new neighbors it has to decide whether it can maintain paths to intended destinations through them.  




\end{itemize}

\begin{table*}
\begin{center}
\caption{Comparison of different architectures and their featutes.}
\label{comparison}
\begin{tabular}{|l||l|l|l|l|}\hline
\em Architecture & \em QoS Support & \em Focus & \em Features & \em Remarks\\\hline \hline
\em JAVeLAN \cite{redi} & No & Low Energy & Dual Rate Transceivers & Redesign of Protocol Stack\\\hline
\em QE$^2$W \cite{lettieri} & Yes & Energy Efficiency & & \\\hline
\em WSNARP \cite{dsouza} & Yes & Remote Power Charging & & Interactive Applications\\\hline
\em uRbAn & Yes & Multipath Routing & Energy Management & Modularity\\
        &    &  & Mobility Management & \\\hline
\end{tabular}
\end{center}
\end{table*}

Table \ref{comparison} shows a comparison between our architecture and the architectures proposed in literature. The architecture proposed in \cite{redi} is named as JAVALeN and it primarily focuses on low energy consumption. However, the architecture does not have a provision of QoS. On the other hand, a {\em QoS-aware energy-efficient wireless node architecture} is proposed in \cite{lettieri}, and a {\em wireless sensor node architecture using remote power charging} is proposed in \cite{dsouza}.  We abbreviate the architecture proposed in \cite{lettieri} as QE$^{2}$W and that proposed in \cite{dsouza} as WSNARPC for reference purposes. Both of these architectures have a provision of QoS. 

In this paper, we propose an architecture of a wireless node that can be used to form an ad hoc network and that has a functionalities to support QoS. Our architecture is modular and is equipped with modules for energy management and mobility management. In the architecture proposed in this paper, the source may use multiple paths to the destination provided that the it is able to identify multiple node-disjoint paths to the destination that satisfy QoS. We would like to mention that identification of multiple node-disjoint paths suffers from path diminution \cite{abbas-jain} and so is true for identification of node-disjoint paths that satisfy the QoS. A survey of methodologies used for providing QoS in ad hoc networks is presented in \cite{abbas-survey}, and the use of multipath routing in QoS provisioning has been discussed in \cite{abbas-ngmast}, \cite{abbas-ndt}, \cite{abbas-jdim}.  

In what follows, we discuss the design of the components that are responsible to provide major functionalities in the proposed architecture.


\section{Routing}
\begin{figure}
\centerline{\psfig{figure=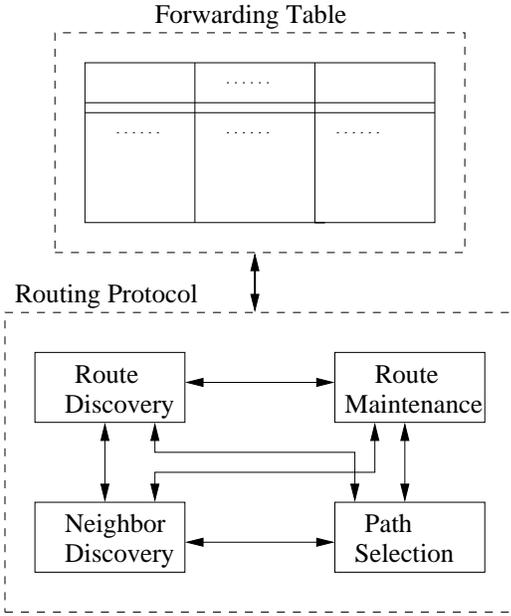, width=3.0in}}
\caption{Functions in routing and their interactions.}
\label{routing}
\end{figure}

The routing component of the node consists of a routing protocol. The routing protocol has four major components: neighbor discovery, route discovery, route maintenance, and path selection mechanism. Figure \ref{routing} shows the routing module functions and their interactions. In what follows, we describe these functions.

\subsection{Neighbor Discovery} 
In the neighbor discovery phase, a node tries to discover who are nodes in its vicinity (or its neighbors). For that purpose, a node periodically transmits {\em hello} packets. The hopcount of a {\em hello} packet is equal to one so that it is not retransmitted by its neighbors. A node that receives a {\em hello} packet sends a reply to the originator of {\em hello} packet to inform that it is alive. After a node receives a reply from a node, it either refreshes its membership or makes a corresponding entry in the list of its neighbors if it was not there previously. If a node does not hear from any of its neighbors for more than a threshold time, it deletes its entry from the list of its neighbors. 

\subsection{Route Discovery}
When a node does not have route to an intended destination, it initiates a route discovery by sending a {\em route request} (RREQ) packet to its neighbors. A node that is neither the source nor the destination of an RREQ is called an intermediate node. When an intermediate node receives a copy of an RREQ, it examines whether its own address is present on the {\em traversed hop list} of the RREQ. If yes, it discards the RREQ. Otherwise, it appends its own address onto the {\em traversed hop list} of the RREQ, and forwards the RREQ to its neighbors according to a stated RREQ forwarding policy. Eventually, the RREQ reaches the destination. The destination is responsible for sending a route reply (RREP). Depending whether a source wishes to have a single path or multiple paths the RREQ forwarding policy and replying procedure at the destination may be customized. 

\subsection{Route Maintenance}
In the route maintenance phase, if a node along a path detects a link failure it tries to repair it through its neighbors. The segment of the path constructed during the repair of the path can only be augmented to the path if it is able to satisfy the QoS requirements of the application. If the repair for any of the current paths passing through cannot be performed, it then informs to the upstream nodes by sending them a {\em route error} (RERR) message. A node receiving an RERR unicasts it to upstream nodes and deletes the entry corresponding to the failed path from its routing table. When an RERR message reaches to the source node, it looks for an alternate path, if any, that satisfies the QoS requirements of the application. If no such path is available and there are packets to be sent to the destination, the source then initiates a route discovery.

\subsection{Path Selection}
For QoS provisioning, paths that satisfy the QoS requirements should be selected out of paths that are available from a source to a destination. Paths may be selected either by the source or by the destination. Alternatively, an RREQ may include the threshold value of the QoS parameter to enable the routing protocol to return only those paths that satisfy the QoS required by the application.

When paths are identified, a node has to decide about the interface to which the packets are forwarded. To do so, a node builds a forwarding table based on the paths identified using the routing protocol. Generally, a forwarding table contains information such as previous hop, interface identifier, and next hop. Note that for all these functions, routing protocol needs to interact with another major function of the node and that is the mobility management. 

\begin{figure}
\centerline{\psfig{figure=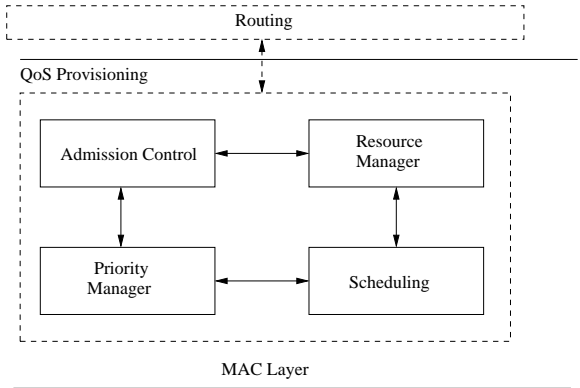, width=3.25in}}
\caption{Functions in QoS provisioning and their interactions.}
\label{qos}
\end{figure}

\begin{figure}
\centerline{\psfig{figure=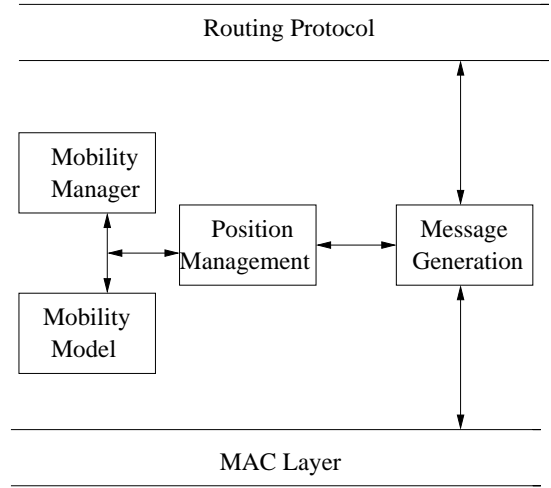, width=3.25in}}
\caption{Components of mobility management and their interaction.}
\label{mobility}
\end{figure}

\section{Mobility Management}
Figure \ref{mobility} shows the components of mobility management and their interactions. There are four major components of the mobility management: {\em mobility manager}, {\em mobility model}, {\em position management}, and {\em message generation}. The {\em mobility manager} is responsible for selecting the parameters such as the destination, speed, direction in accordance with the mobility model selected. The {\em mobility model} contains what is the model to be followed by the node for moving from its current position. There can be different mobility model, however, a node has to select and follow only one mobility model at a time. Although, a node can select any model from a list of available mobility models, however, the selection of the mobility model, in general, is carried out in accordance with the decisions of the participating nodes in the network. The {\em position management} component is responsible for managing the current position of the node. For that it may need to interact with some geographic or global positioning system as well. The {\em message generation/reception} component is responsible to generate and receive messages to and from routing and MAC layer protocols. Note that mobility management can be a part of either the operating system or it may be a firmware depending upon what mobility pattern is followed in a specific application. For example, in a vehicular ad hoc network, road network is required for the mobility. 

In what follows, we describe yet another function and that is the QoS provisioning.
 
\section{QoS provisioning}
The QoS provisioning part includes the functions such as admission control, priority manager, resource manager, and scheduling. We describe these functions as follows. 

\subsection{Admission Control}
In the {\em admission control}, a node has to determine what may be the consequences of allowing a packet into the network. If allowing the packet is not detrimental to the performance of the network, the packet is allowed to enter into the network. Otherwise, the packet is discarded. This is done to ensure that the ongoing flows of packets or connections should not suffer due to the packets that have not yet entered into the network. The packets that were discarded may enter into the network at a later time provided that enough resources are available. 

\subsection{Priority Manager}
The {\em priority manager} decides what should be the priority of the packet so that it is likely to reach the destination on or before the deadline. It may increase the priority of the packet if it deems that the packet needs to be expedited so as to catch the deadline. However, it depends upon how many packets of a particular priority level have been forwarded during the current time window as it affects the service received by a class of packets. Conversely, it may decrease the priority of the packet if the packet has arrived too early. In other words, the priority of the packet may depend upon the level of urgency of the packet and service received by the packet upstream. 

\subsection{Resource Manager}
The {\em resource manager} is responsible for managing the resources so as to provide the QoS desired by the application. It reserves the resources needed by the flow of packets, and it releases the resources when they are not needed. The resources can be reserved apriori for an ensuing packet transmission or can be reserved ondemand. However, only one type of resource reservation scheme from the available schemes can be selected at a time.   

\subsection{Scheduling}
The {\em scheduling} component schedules the packet to be forwarded by the node onto the output link. Note that QoS provisioning module interacts with mobility management module to handle the mobility of nodes. Also, it interacts with routing and MAC layer for the effective provision of QoS.  Figure \ref{qos} shows various functions and how they interact for the provision of QoS.

Note that energy is a scarce resource in case of mobile ad hoc network. In what follows, we describe the energy management module of the wireless node.

\begin{figure}
\centerline{\psfig{figure=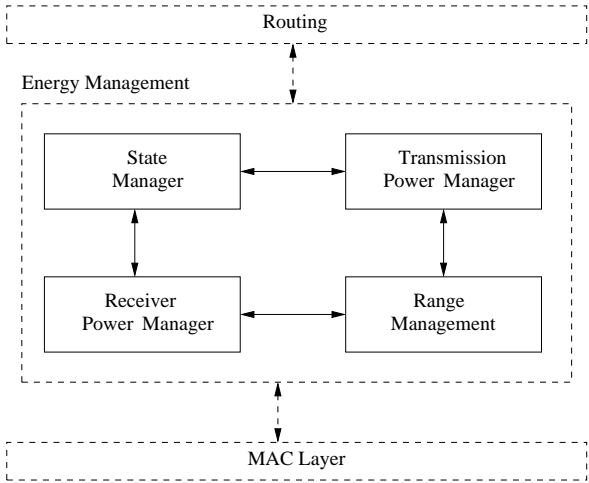, width=3.25in}}
\caption{Functions in energy management and their interactions.}
\label{energy-management}
\end{figure}

\begin{figure}
\centerline{\psfig{figure=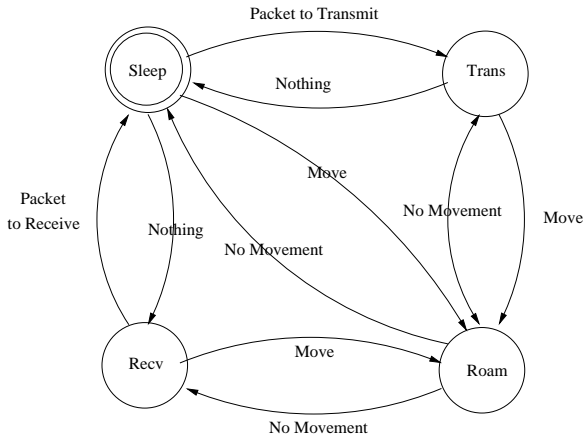, width=3.25in}}
\caption{State transition diagram for energy management.}
\label{state-transitions}
\end{figure}


\section{Energy Management}
The energy management module consists of state manager, transmission power manager, receiver power manager, and range management. The {\em state manager} is responsible at what time the node will switch its states. A node can be in the following states: sleep, transmit, receive, roaming. Figure \ref{state-transitions} shows a state transition diagram for energy management. The sleep state is the initial state of the node. If the node has packets to transmit, it enters into a the transmit state, and after finishing the transmission, if it has nothing to do (i.e. it does not have packets to receive and it is not moving) then it goes back to the sleep state.  In other words, a node which is not either in {\em transmit} or {\em receive} or {\em roaming} states for a timeout may enter into a {\em sleep} state. A node may enter to a roaming state from a transmit/receive/receive when it starts to move and vice versa. When a node moves from its position it enters into the roaming state meaning that it needs to update its position if it is equipped with a global positioning system (GPS) and takes other necessary actions. Note that in the sleep state, the node is supposed to spend minimum power. In the transmit state the power spent is higher as compared to that spent in the receive mode. The {\em transmitter power manager} manages the power required to transmit the packet. Similarly, the {\em receiver power manager} is responsible for managing the power of the receiver. The purpose of the {\em range management} module is how much transmission range/receiver range is required to be adjusted so as to correctly transmit/receive the packets.

\section{Miscellaneous Functions}
There are miscellaneous components such as device management, memory and processor management. Actually these functions are taken care of by the operating system. Generally, the {\em interface manager} and {\em device manager} are also part of the operating system as it manages and controls the hardware devices. We defer the design and details of these functions as it is not the appropriate place to address them. The readers are referred to \cite{silberschatz} for functions related to the operating system. There are some other components that are built into the operating system itself. For example, transport protocols such as {\em transmission control protocol} (TCP) and {\em user datagram protocol} (UDP) and their variations for a wireless node may be incorporated into the operating system itself.  All these functions can be expanded following a procedure as that in previous subsections.   

Note that our design is modular. Also, 
it may be customized to fit into an SoC or SiP. In what follows, we conclude the paper.

\section{Conclusion}
The design of a wireless node that can be used to form an ad hoc network with QoS support is a challenging task. In this paper, we proposed an architecture for such a node. In our architecture, we first identified the major functionalities and then those functionalities are divided into subtasks to be carried out in an iterative manner. A special feature of our architecture is its ability to utilize multiple paths, if any, for the provision of QoS. However, it can also work with a single path routing in case a multipath routing protocol is not employed.  In our design, we followed a modular approach. Our design 
may be customized to fit either into an SoC or an SiP. Further validation forms the future work.


\begin{thebibliography}{99}

\bibitem{wang} Y. Wang, D. Zhao, ``Design and Implementation of Routing Scheme for Wireless Network on Chip'', {\it Proceedings of IEEE International Symposium on Circuits and Systems (ISCAS)}, pp. 1357-1360, May 2007.

\bibitem{zhao} D. Zhao, Y. Wang, H. Wu, ``Dual-Channel Binary-Countdown Medium Access Control in Wireless Network-on-Chip'', {\it Proceedings of 2nd ACM International Conference on Nano-Networks (NanoNet)}, Article 11, pp. 1-5, September 2007. 

\bibitem{kim} M. Kim, D. Kim, G.E. Sobelman, ``Network-on-Chip Quality of Service Through MultiProtocol Label Switching'', {\it Proceedings of IEEE International Symposium on Circuits and Systems (ISCAS)}, pp. 1843-1846, September 2007.

\bibitem{redi} J. Redi, J. Kolek, K. Manning, C. Patridge, R.R. Hain, R. Ramanathan, I. Castineyra ``JAVALeN: An Ultra-Low Energy Ad hoc Wireless Network'', {\it Elsevier Journal on Ad Hoc Networks}, Vol. 6, pp. 108-126, September 2008. 

\bibitem{cidon} I. Cidon, I. Keidar, ``Zooming in on Network-on-Chip Architectures'', Technical Report CCIT 565, Technion Department of Electrical Engineering, December 2005.

\bibitem{secchi} S. Secchi, F. Palumbo, D. Pani, L. Raffo, ``A Network on Chip Architecture for Heterogeneous Traffic Support with Non-Exclusive Dual-Mode Switching'', {\it 11th EUROMICRO Conference on Digital System Design Architectures, Methods and Tools (DSD)},  pp. 141-148, September 2008.

\bibitem{kishi} Y. Kishi,  K. Tabata, T. Kitahara, Y. Noishiki, A. Idoue, S. Nomoto, ``Wireless Node Architecture and Its Implementation for Multi-Hop Mesh Networks in IP-Based Broadband Fixed Wireless Access Systems'', {\it IEICE Transactions on Communications},  Vol. E88B, No. 3, pp. 1202-1210, 2005.

\bibitem{lettieri} P. Lettieri,  M.B. Srivastava, ``A QoS-Aware, Energy-Efficient Wireless Node Architecture'', {\it Proceedings of IEEE International Workshop on Mobile Multimedia Communications (MoMuC)}, pp. 252-261, November 1999.

\bibitem{dsouza} M. D'Souza, K. Bialkowski, A. Postula, M. Ros, ``A Wireless Sensor Node Architecture Using Remote Power Charging, for Interaction Applications'', {\it Proceedings of the 10th Euromicro Conference on Digital System Design Architectures, Methods and Tools (DSD)}, pp. 485-494, September 2007. 

\bibitem{silberschatz} A. Silberschatz, P.B. Galvin, G. Gagne, Concepts of Operating Systems, John Wiley and Sons, Seventh Edition, 2006.

\bibitem{ji} M. Ji, Y. Kim, M. Seo, J. Ma, ``A Novel Multi-Channel Multi-Radio Wireless Mesh Node Architecture for NS-2'', {\it Proceedings of IEEE International Conference on Communications and Mobile Computing (MCM)} Vol. 2, pp. 147-151, January 2009.


\bibitem{abbas-survey} A.M. Abbas, O. Kure, ``Quality of Service in Mobile Ad hoc Networks: A Survey'', {\it International Journal of Ad hoc and Ubiquitous Computing (IJAHUC)}, vol. 6, no. 2, pp. 75-98, June 2010.

\bibitem{abbas-ngmast} A.M. Abbas, O. Kure, ``CQSR: A Correlation Aware Quality of Service Routing in Mobile Ad Hoc Networks'', {\em Proceedings of 3rd IEEE International Conference and Exhibition on Next Generation Mobile Applications, Services and Technologies (NGMAST)}, pp. 363-368, September 2009. 

\bibitem{abbas-ndt} A.M. Abbas, O. Kure, ``A Probabilistic Quality of Service Routing in Mobile Ad Hoc Networks'', {\em Proceedings of 1st IEEE International Conference on Networked Technologies (NDT)}, pp. 269-273, July 2009.

\bibitem{abbas-jdim} A.M. Abbas, O. Kure, ``A Deadline-Driven Probabilistic Quality of Service Routing for Mobile Ad hoc Networks'', {\em Journal of Digital Information Management (JDIM)}, vol. 8, no. 2, pp. 136-142, April 2010. 

\bibitem{abbas-jain} A.M. Abbas, B.N. Jain, ``Path Diminution is Unavoidable in Node-Disjoint Multipath Routing for Mobile Ad hoc Networks with Single Route Discovery'', {\it International Journal of Ad hoc and Ubiquitous Computing (IJAHUC)}, vol. 5, no. 1, pp. 7-21, January 2010.



\end{thebibliography}
\end{document}